\documentclass[a4paper]{jpconf}
\usepackage{graphicx}
\begin{document}
\title{Fractional Action Cosmology with Power Law Weight Function}

\author{Mubasher Jamil}

\address{Center for Advanced
Mathematics and Physics, National University of Sciences and
Technology, H-12, Islamabad, Pakistan}

\ead{mjamil@camp.nust.edu.pk}

\author{Muneer A. Rashid}

\address{Center for Advanced
Mathematics and Physics, National University of Sciences and
Technology, H-12, Islamabad, Pakistan}

\ead{muneerrshd@yahoo.com}

\author{D. Momeni}

\address{Department of Physics, Faculty of Sciences, Tarbiat Moa'llem University, Tehran,
Iran}

\ead{d.momeni@yahoo.com}

\author{Olga Razina}

\address{Eurasian International Center for Theoretical Physics,
Eurasian National University, Astana 010008, Kazakhstan}

\author{Kuralay Esmakhanova}

\address{Eurasian International Center for Theoretical Physics,
Eurasian National University, Astana 010008, Kazakhstan}

\begin{abstract}
Motivated by an earlier work on fractional-action cosmology with a
periodic weight function \cite{nabulsi}, we extend it by choosing a
power-law weight function in the action. In this approach, we obtain
a varying gravitational coupling constant. We then model dark energy
in this paradigm and obtain relevant cosmological parameters.
\end{abstract}

\section{Dirac's Large Numbers Hypothesis and time running of $G$}

During our survey of various physical laws of nature, we came across
a number of constants entangled with these laws. Historically it was
Weyl \cite{Weyl1917,Weyl1919} who initiated the idea of large
numbers. However, it was Dirac who discovered an apparently unseen
thread joining up these physical constants by a simple yet
interesting law, viz., \emph{Law of Large Numbers}. Using this law,
Dirac arrived at his \emph{Large Number Hypothesis} (LNH) which had
profound influence on the world of physics. It generated a lot of
activity both at theoretical and observational level with LNH as
their starting point. Some works are centered around modification of
Einstein's gravitational theory and related equations to adapt the
$G$ variation \cite{Dirac1973,Canuto}. Some authors have also given
counter-arguments against LNH for justifying and refuting that
hypothesis characterize the second category of works in favor of it
\cite{Bousso} and also testing the validity of it \cite{Blake}. Even
there is found a formal link between LNH and cosmological constant
term $\Lambda$ \cite{Peebles}. Weyl's large number is square root of
Eddington's number $N\sim10^{80}$ (which represents the total number
of protons in the Universe) \cite{Eddington} and is supported by
observational data \cite{Stewart}. Indeed Stewart showed  that the
ratio of the radius of the Universe and the electron is only two
orders of magnitude smaller $10^{40}$ than that of Weyl's number.
There are some other large numbers like Jordan's number
\cite{Jordan} and the Shemi-zadeh's number \cite{Shemi-zadeh} which
have also appeared in the literature.

As originally stated by Dirac there are three dimensionless numbers
in nature which can be constructed from the atomic and cosmological
data:
\begin{enumerate}
  \item The ratio of the electric to the gravitational force between an
electron and a proton $\frac{e^2}{Gm_em_p}\simeq7\times10^{39}$.
\item The mass of that part of the Universe that is receding from
us with a velocity $v<c/2$ expressed in units of the proton mass is
$10^{78}$.
  \item The age of the Universe $t$, expressed in terms of a unit of time
   $\frac{e^{2}}{m_{e}c^{3}}$
provided by atomic constants, is inversely proportional to Newton's
gravitational constant.
\end{enumerate}
The above three statements imply in respective order
\begin{eqnarray}
\frac{e^{2}}{Gm_{e}m_{p}}\propto t,\\
M\propto t^{2},\\
G\propto t^{-1}.
\end{eqnarray}
There is a serious problem between (3) and GR
\cite{Dirac1973,Dirac}: Einstein's theory requires $G$ to be a
constant. As was noted by Dirac, this inconsistency might be
resolved if\emph{``.. we assume that Einstein theory is valid in a
different system of units from those provided by the atomic
constants.''} Consequence of Dirac's LNH is the existence of a
variable $G$ cosmology. A beautiful account of the constants of
nature appears in \cite{barrow}.

\section{Framework}

Recently Nabulsi \cite{nabu} examined the Friedmann-Robertson-Walker
(FRW) spacetime in the context of fractional action-like variational
approach or fractionally differentiated Lagrangian function
depending on a fractality parameter $\chi$ to model nonconservative
and weak dissipative classical and quantum dynamical systems.

We apply the technical machinery of differential geometry to general
relativity in the context of fractional action-like variational
approach. Following Einstein's arguments, we start by representing
the gravitational field by an affine connection on a curved manifold
and assume that the free fall corresponds to geodesic motion. Thus
the classical Lagrangian of a curve is defined by $L=(1/2)g_{ij}\dot
x^i\dot x^j$, where $\dot x^i=dx^i/dt$ and $g_{ij}$ is the metric
tensor. This gives the action
%%%
\begin{eqnarray}
S=\frac{m}{2}\int\limits_0^\tau
g_{ij}(x)\dot{x}^{i}\dot{x}^{j}\mu(\chi,t)dt,\label{action1}
\end{eqnarray}
%%%%
where $m$ is the particle's mass. Here $\mu(\chi,t)$ is the
generalized weight function and if we choose it to be a periodic
function, we recover the results in \cite{nabulsi}. This weight
function can produce a friction force in classical dynamics for the
test particle and as we will show, it also yields a time varying
$G$. The modified resultant fractional Euler-Lagrange equations are
%%%
\begin{eqnarray}
 \ddot{x}^{l}+\partial_{t}\log(\mu(\chi,t))\dot{x}^{l}+
\Gamma^{l}_{ij}\dot{x}^{i}\dot{x}^{j}=0.
\end{eqnarray}

%%%%
We follow the procedure of \cite{nabu} and assume that the test
particle's velocity is approximately zero. Furthermore, the metric
and its derivative are approximately static. Applying these
 assumptions to the spatial components of (5) gives:
%%%

\begin{eqnarray}
 \partial_{i}\overrightarrow{\gamma
}^{i}+\partial_{t}\mu\partial_{i}v^{i}+R_{tt}=0.
 \end{eqnarray}
%%%%

In the above equation $v^i=dx^i/dt$ and $\gamma^i=dv^i/dt$ are the
particle's velocity and acceleration respectively and
$\partial_{i}\gamma ^{i}=-4\pi G\rho$, which is Poisson's equation
in Newtonian gravity. Here $\rho$ is the mass density in
gravitational interaction and $G$ is Newton's gravitational
constant. The resultant modified
 gravitational constant is

%%%
\begin{eqnarray}
 G'=G\Big(1-\frac{\partial_{t}\log(\mu(\chi,t))}{4\pi
G\rho}\partial_{i}v^{i}\Big)=G+\triangle G.
\end{eqnarray}
%%%%

 If $\mu(\chi,t)=t$ then (7) describes Dirac's
 model \cite{Dirac}. Thus the change in $G$ in a spatially flat
 $(D+2)$-dimensional FRW Universe is
$$
\triangle G=-\frac{D(D+1)H\partial_{t}\log(\mu(\chi,t))}{8\pi\rho},
$$
where $H=\partial_{t}(\log(a(t))$ is the Hubble parameter. The
Friedmann equation for such a model with metric
$g_{ij}=diag(1,-a^{2}(t)\eta_{AB})$, $A,B=2,...,D+2,$ with
$\eta_{AB}=diag(1,...,1)$ is

%%%
\begin{eqnarray}
 \frac{D(D+1)}{2}H^2=8\pi G\rho+\Lambda.
 \end{eqnarray}
%%%%
One thinks that if the gravitational constant changes slightly, as
we have expressed it in this work, the Friedmann's equation must be
replaced by an effective $G$. A simple example of effective $G$ is
the $f(R)$ gravity. An interesting feature of these theories is the
fact that the gravitational constant is both time and scale
dependent. After some calculations, one can obtain a Poisson
equation in the Fourier space and attribute the extra terms as
effective gravitational constant $G_{eff}$ \cite{f(R)}. But now, we
are working using an Einstein-Hilbert action therefore there is no
need to insert effective $G$ on the right hand side of the
Friedmann's equation (8).

%%%%%%%%%%%%%%%%

 Using the Bianchi identity and assuming both
energy density $\rho$ and the cosmological constant $\Lambda$ as
functions of time related by $\Lambda(t)=\Lambda_{0}\rho(t)^n$.
Consequently we have
%%%
\begin{eqnarray}
 G(t)=\frac{G}{\mu(\chi,t)} ,\ \ \
\Lambda(t)=\Lambda_{n}\mu(\chi,t)^{n/(1-n)} ,\ \ \
\rho(t)=\rho_n\mu(\chi,t)^{1/(1-n)},
\end{eqnarray}
%%%

where $ \Lambda_{n}\equiv\Lambda_{0}(\frac{8\pi G(1-n)}{\Lambda_{0}
n })^{n/(n-1)}$ and $\rho_n\equiv (\frac{8\pi G(1-n)}{\Lambda_{0}
n})^{1/(n-1)}$. Substituting (9) in (8) we obtain
%%%
\begin{eqnarray}
 H=\frac{\dot a}{a}=c_{n}\mu(\chi,t)^{n/(2(1-n))},
 \end{eqnarray}

%%%

where $c_{n}=\sqrt{\frac{2b_{n}}{D(D+1)}}$ and $b_{n}=8\pi
G\rho_n+\Lambda_n$  and $0<n<1$. By solving (10), we find
%%%

\begin{eqnarray}
 a(t)=a_{0}\exp\Big(c_{n}\int \mu(\chi,t)^{n/(2(1-n))}dt\Big).
 \end{eqnarray}

%%%%
%%%%%%%%%%%%%%%%%%%%%%%%%%%%
For the remaining part of the paper, we use power law weight
function as
%%%
\begin{eqnarray}
 \mu(\chi,t)=\chi t^{p},\ \ \ \ p\geq1.
  \end{eqnarray}
  %%%
 One can treat (12) as a generalized Dirac's hypothesis.
  Note that $p=1$ leads to an
 inverse power of time relation for the gravitational constant
  i.e. $G= \frac{G_0}{t}$. Substituting (12) in
 (11) yields
 %%%
 \begin{eqnarray}
 a(t)=a_{0}\exp(\alpha t^{\beta}),
\end{eqnarray}
 %%%
where
\begin{equation}
\alpha\equiv\chi^{\frac{n}{2(1-n)}}\frac{2c_{n}(1-n)}{2+(p-2)n},\ \
\beta\equiv\frac{2+(p-2)n}{2(1-n)}.\nonumber
\end{equation}
It is useful to discuss the behavior of the scale factor
diagrammatically. For simplicity we take $a_0=1$ and $\alpha=-1$,
and focuss only on Dirac's model i.e. $G(t)=\frac{G}{t}$. Figure 1
shows the behavior of the scale factor as a function of time for
different values of $n$.
\begin{figure}
\centering
 \includegraphics[scale=0.5]{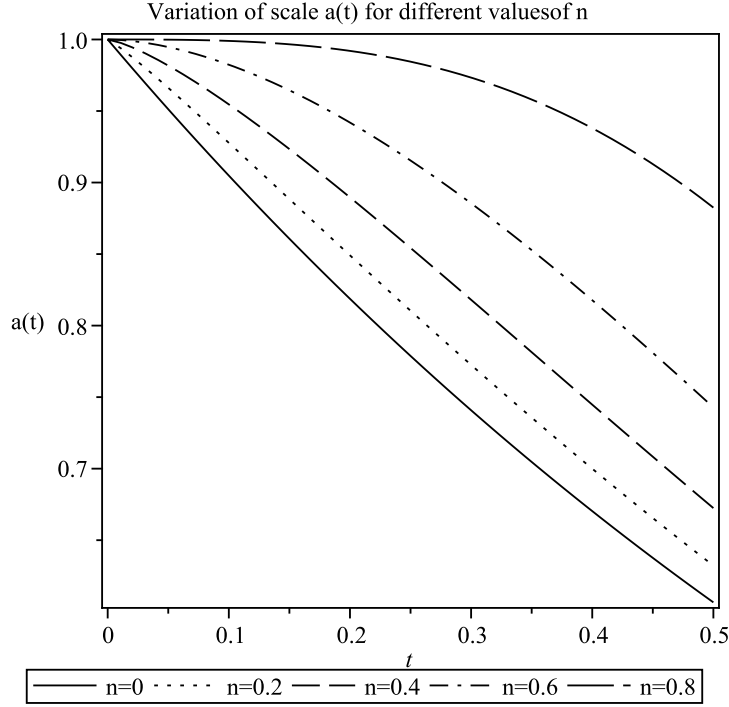} % scale goes from 0 to 1.
 \caption{ Variation of scale factor (13) for different values of $n$.}
  \label{fig1.eps}
\end{figure}

The Hubble parameter is
%%%
\begin{eqnarray}
 H=\alpha\beta t^{\beta-1}=\alpha\beta\Big(-\frac{1}{\alpha}\ln(1+z)\Big)^{1-\frac{1}{\beta}}.
 \end{eqnarray}
%%%
Here redshift $z$ is defined by $1+z=\frac{a_0}{a(t)}$. Figure 2
shows the behavior of the Hubble parameter as a function of time for
$\alpha=1$.

\begin{figure}
\centering
 \includegraphics[scale=0.5]{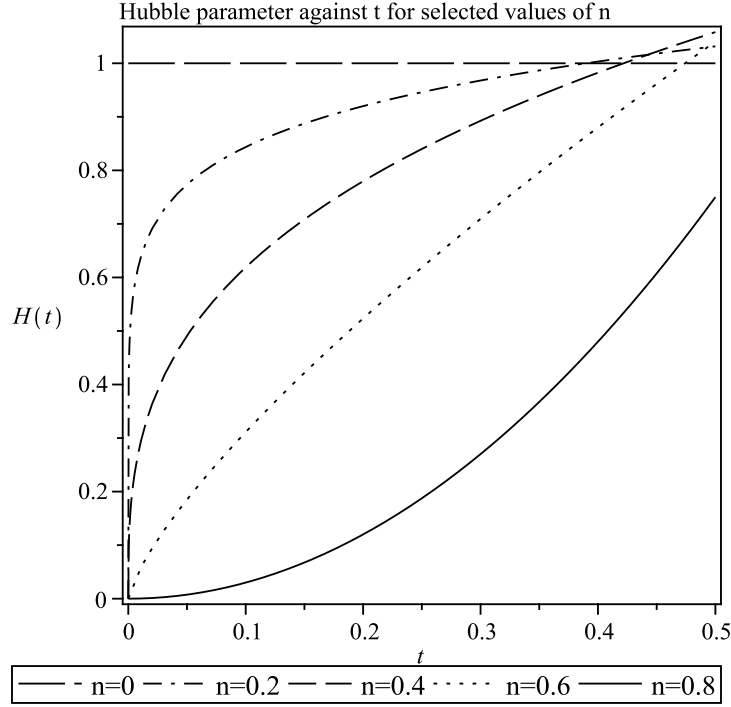} % scale goes from 0 to 1.
  \caption{ Hubble parameter against $t$ for selected values of $n$.}
  \label{fig2.eps}
\end{figure}

We note that the scale factor for $n=\frac{2}{3}$ ,
$\alpha=\frac{1}{2}$ is $a\propto \exp(\frac{t^2}{2})$.
Consequently, the scale factor would increase very rapidly for the
positive branch, while it would start from infinity for the negative
branch. Also the choice $n=\frac{2}{3}$ , $\alpha=\frac{-1}{2}$
gives $a\propto \exp(\frac{-t^2}{2})$. This indicates that the
Universe is evolving with a Hubble parameter $H\sim -t$ and a scale
factor $a\propto \exp(\frac{-t^2}{2})$, which is close to
singularity related to the origin and the fate of the Universe. The
same phenomena occurs in the context of a new model of dark energy,
the Quintom model \cite{Quintom}.

\subsection{Deceleration parameter}

 We know that
%%%
\begin{eqnarray}
 q = -1-\frac{\dot{H}}{H^2},
  \end{eqnarray}
%%%

is the deceleration parameter. For an expanding Universe, $q>-1$.
Using (14) we can write (16) as
%%%
\begin{eqnarray}
 q=-1-\frac{1-\beta}{\beta}\frac{1}{\ln(1+z)}.
  \end{eqnarray}
%%%
 For Dirac's hypothesis
($p=1$), we have
%%%
\begin{eqnarray}
 q=-1+\frac{n}{2-n}\frac{1}{\ln(1+z)}.
 \end{eqnarray}
%%%
 Thus we conclude that the Universe can  remain in an accelerating
  expansion for some values of $z$.
The next diagram (Figure 3) represents the graph of $q$ versus $z$
for different values of $n$. It shows that when $n$ increases the
deceleration parameter $q$ recedes from the crossing line $q=-1$
\footnote{remembering that the $0<n<1$}. Thus we can model cosmic
acceleration in the present framework.
\begin{figure}
\centering
\includegraphics[scale=0.5]{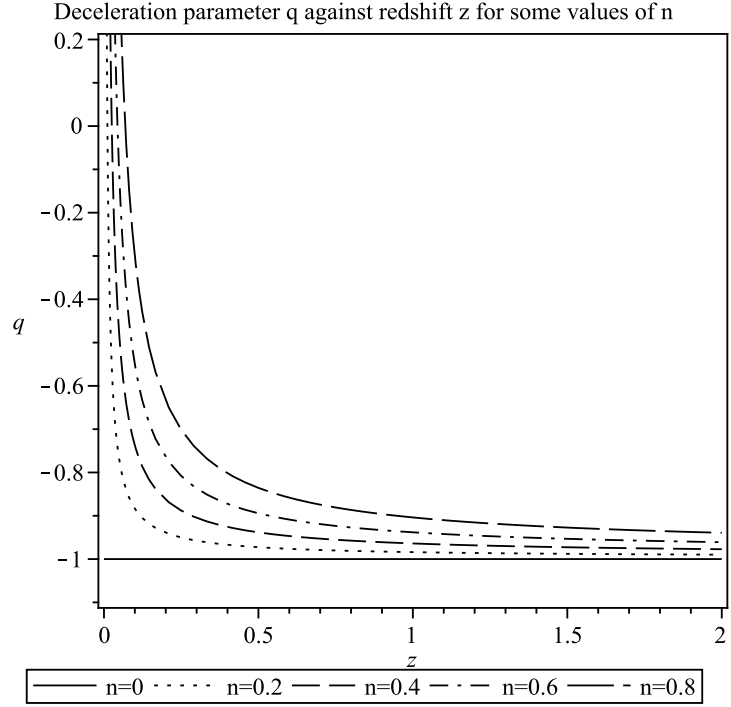} % scale goes from 0 to 1.
 \caption{ Deceleration parameter $q$ against redshift $z$ for some values of $n$.}
  \label{fig3.eps}
\end{figure}
%%%%%%%

\subsection{The statefinder parameters for dark energy}

The statefinder parameters $\{r,s\}$ were introduced to characterize
primarily the spatially flat Universe ($k = 0$) models with cold
dark matter (dust) and dark energy \cite{sahni}. They were defined
as
%%%
\begin{eqnarray}
 r\equiv\frac{\ddot{a}}{a H^3}={\frac { \left(
{\beta}^{2}-3\,\beta+2 \right) {t}^{-2\,\beta}+\beta\,
 \left(  \left( 3\,\beta-3 \right) {t}^{-\beta}+\alpha\,\beta \right)
\alpha}{{\alpha}^{2}{\beta}^{2}}},
\end{eqnarray}

%%%
\begin{eqnarray}
 s\equiv\frac{r-1}{3(q-\frac{1}{2})}=-\frac{2}{3}\,{\frac {
\left( \beta-1 \right)  \left( {t}^{-\beta}\beta-2\,{t}
^{-\beta}+3\,\alpha\,\beta \right) }{\alpha\,\beta\, \left(
3\,\alpha \,{t}^{\beta}\beta+2\,\beta-2 \right) }}.
\end{eqnarray}

%%%
From (19) and (20) we obtain the relation between $r$ and $s$ as
%%%
\begin{eqnarray}
 r={\frac {{\alpha}^{2-\beta} \left(
2+7\,s\beta\,y-4\,\beta-9\,s+16\,s
\beta+2\,{s}^{2}{\beta}^{2}-13\,{s}^{2}\beta-7\,s{\beta}^{2}-7\,sy-2\,
\beta\,y+2\,{\beta}^{2}+11\,{s}^{2}+2\,y \right) }{ \left(
-\beta+1-s \beta+s+y \right) ^{2}}},
 \end{eqnarray}
 %%%

where

%%%
\begin{eqnarray}
 y=\sqrt
{{\beta}^{2}-2\,\beta+2\,s\beta+1-2\,s+{s}^{2}{\beta}^{2}-2\,{
s}^{2}\beta+{s}^{2}}.
 \end{eqnarray}

 %%%

We know that the cosmological constant possesses a fixed equation of
state parameter $(w = -1)$ and a fixed Newton's gravitational
constant, hence $\{1,0\}$ corresponds to $\Lambda$CDM. Moreover
$\{1,1\}$ represents the standard cold dark matter model containing
no radiation while Einstein's static Universe corresponds to
$\{-\infty,\infty\}$ \cite{jamil}.

Note that for Dirac's hypothesis $p=1,n=0,\beta=1$, we have
$\{r,s\}=\{1,0\}$ which represents a static cosmological constant
with cold dark matter. In Figure-4, we draw the pair $\{r,s\}$ for
different values of $\beta=\frac{2-n}{2(1-n)},\alpha=1$. As we can
see, the fixed gravitational constant occurs several times for some
values of $n$. Also our model describes the standard cold dark
matter and a static Einstein Universe for some values of $n$.

\begin{figure}
\centering
 \includegraphics[scale=0.5]{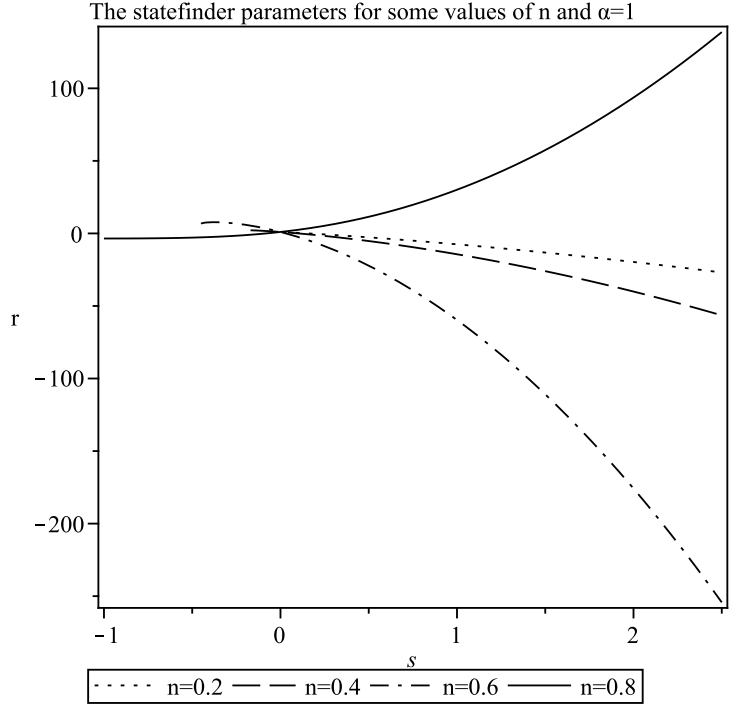} % scale goes from 0 to 1.
  \caption{ The statefinder parameters for some values of $n$.}
  \label{fig4.eps}
\end{figure}

\subsection{Lookback time}
 If a photon is emitted by a source at the
instant $t$ and received at time $t_0$, then the photon travel time
or the lookback time $t- t_0$ is defined by
%%%
\begin{eqnarray}
 t-t_0=-\int_{0}^{z}\frac{dz}{H(z)
(1+z)},
\end{eqnarray} %%%

where $a_0$ is the present value of the scale factor of the
Universe. It is appropriate to calculate this time as a function of
$z$. After substituting Eq. (15) in Eq. (23) we have

%%%
\begin{eqnarray}\nonumber
 t-t_0=-\frac{(-\alpha)^{1-1/\beta}}{\alpha\beta}\int_{0}^{z}\frac{dz}{(\log(1+z))^{1-1/\beta}
(1+z)},
\end{eqnarray} %%%
Now we change the variable as $x=\log(1+z)$. Calculating the
integral above, we find
%%%
\begin{eqnarray}
 t-t_0=(-\frac{\log(1+z)}{\alpha})^{\frac{1}{\beta}}.
   \end{eqnarray}
 Figure-5 shows the behavior of the lookback
time as a function of redshift for values $\alpha=-1,0<n<1$.

\begin{figure}
\centering
 \includegraphics[scale=0.5]{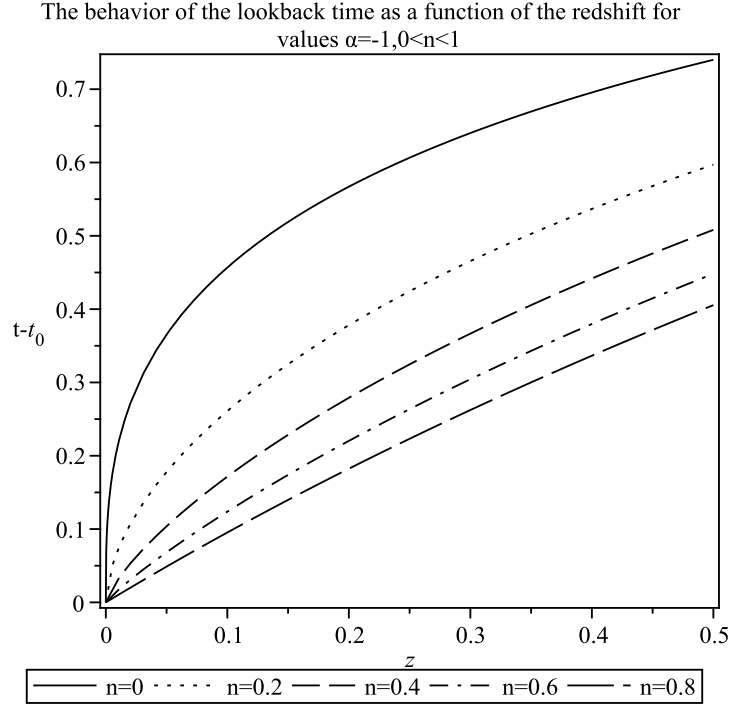} % scale goes from 0 to 1.
  \caption{ The  behavior of the lookback time as a
function of redshift for $\alpha=-1,0<n<1$.}
 \label{fig5.eps}
\end{figure}

\subsection{Proper distance}
If a photon is emitted by a source and is received by an observer at
time $t_0$ then the proper distance between them is defined by
%%%
\begin{eqnarray}
 d=a_0\int_{a}^{a_0}\frac{da}{aH^2}=\int_{0}^{z}\frac{dz}{H^2(z)(1+z)}.
 \end{eqnarray}
%%%
Using Eq. (15) in (25), we get
%%%
\begin{eqnarray}
d=\frac{1}{\beta(\beta-2)}(-\frac{\ln(1+z)}{\alpha})^{-1+\frac{2}{\beta}}.
    \end{eqnarray} %%%
Figure 6 shows the behavior of proper distance in the present model.
\begin{figure}
\centering
 \includegraphics[scale=0.5]{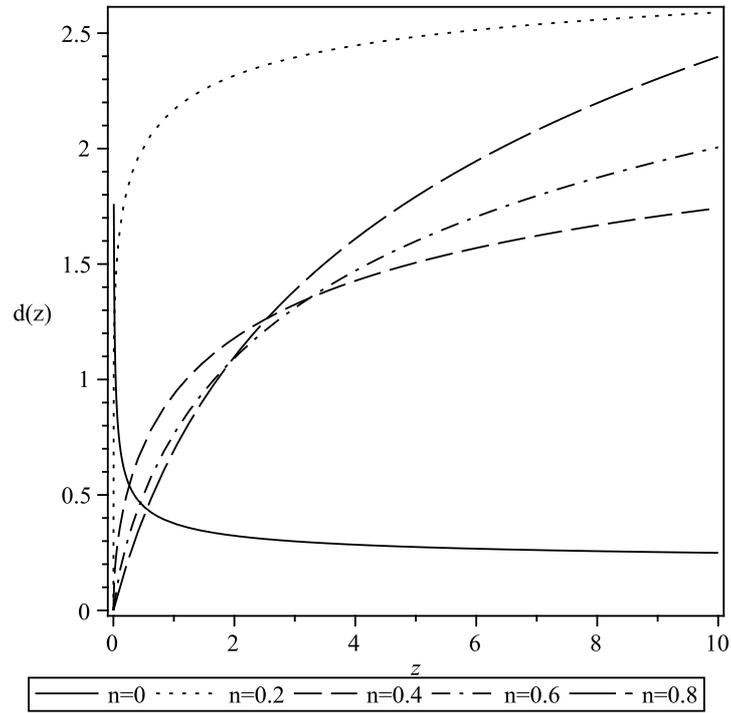} % scale goes from 0 to 1.
  \caption{ The  behavior of the proper distance as a
function of redshift for $\alpha=-1,0<n<1$.}
  \label{fig6.eps}
\end{figure}

\subsection{Luminosity distance}
If $L$ is the total energy emitted by the source per unit time and
$\ell$ is the apparent luminosity of the object, then the luminosity
distance $d_{L}=\left(\frac{L}{4\pi\ell}\right)^{\frac{1}{2}}$
evolves as
%%%
\begin{eqnarray}
d_{L}=d(1+z)=\frac{z+1}{\beta(\beta-2)}\Big(-\frac{\ln(1+z)}{\alpha}\Big)^{-1+\frac{2}{\beta}}.
\end{eqnarray} %%%

Figure (7) shows the variation of
 the luminosity distance as a function of the exponent $0<n<1$.

\begin{figure}
\centering
 \includegraphics[scale=0.5]{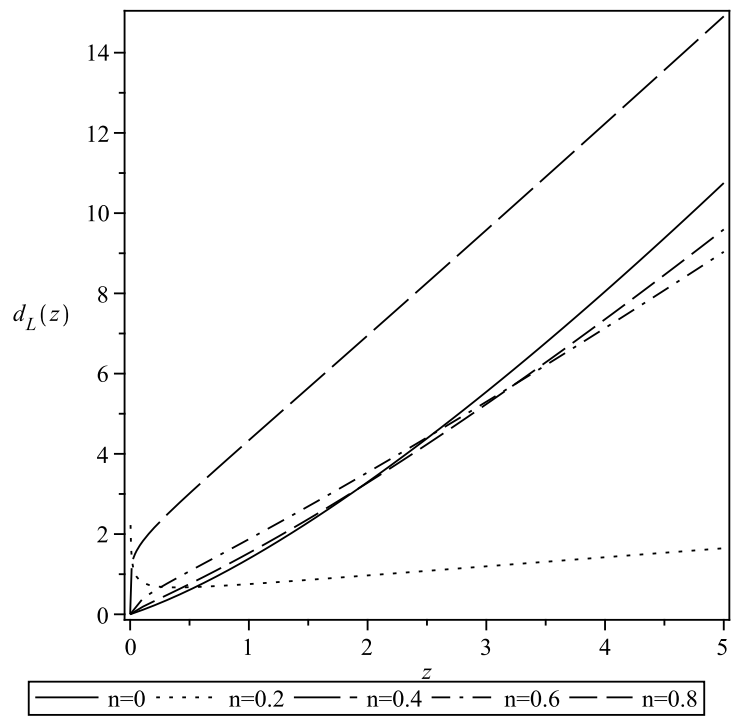} % scale goes from 0 to 1.
  \caption{ The  behavior of the luminosity distance as a
function of redshift for $\alpha=-1,0<n<1$.}
  \label{fig7.eps}
\end{figure}

\subsection{Angular diameter}
The angular diameter distance $d_A$ is
%%%
\begin{eqnarray} d_A=d_L
(1+z)^{-2}=\frac{1}{\beta(\beta-2)(1+z)}(-\frac{\ln(1+z)}{\alpha})^{-1+\frac{2}{\beta}}.
\end{eqnarray}
 %%%

Figure-8 shows the variation of the angular diameter as a function
of the exponent $0<n<1$.
\begin{figure}
\centering
 \includegraphics[scale=0.5]{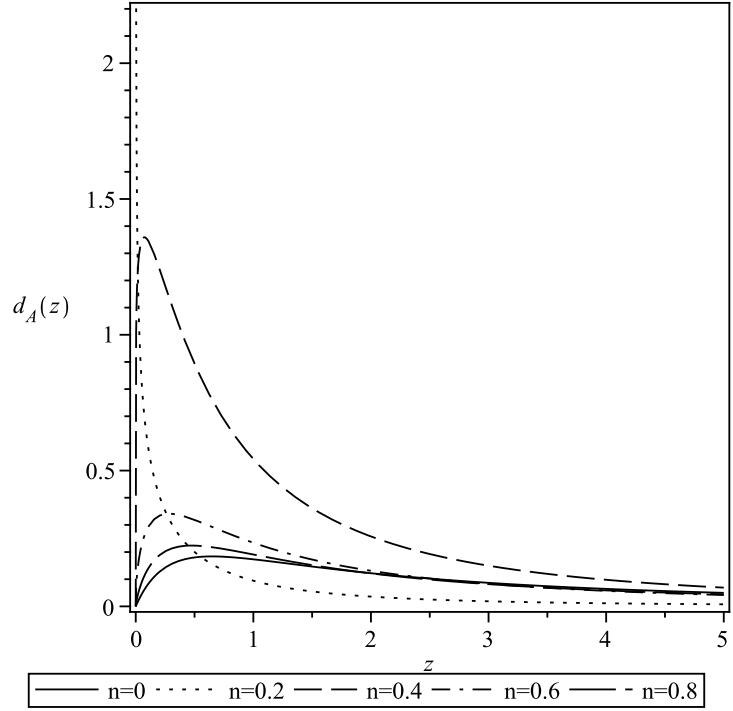} % scale goes from 0 to 1.
  \caption{ The  behavior of the angular diameter as a
function of redshift for $\alpha=-1,0<n<1$.}
  \label{fig8.eps}
\end{figure}

\subsection{State parameter of dark energy}
\begin{figure}
\centering
 \includegraphics[scale=0.5]{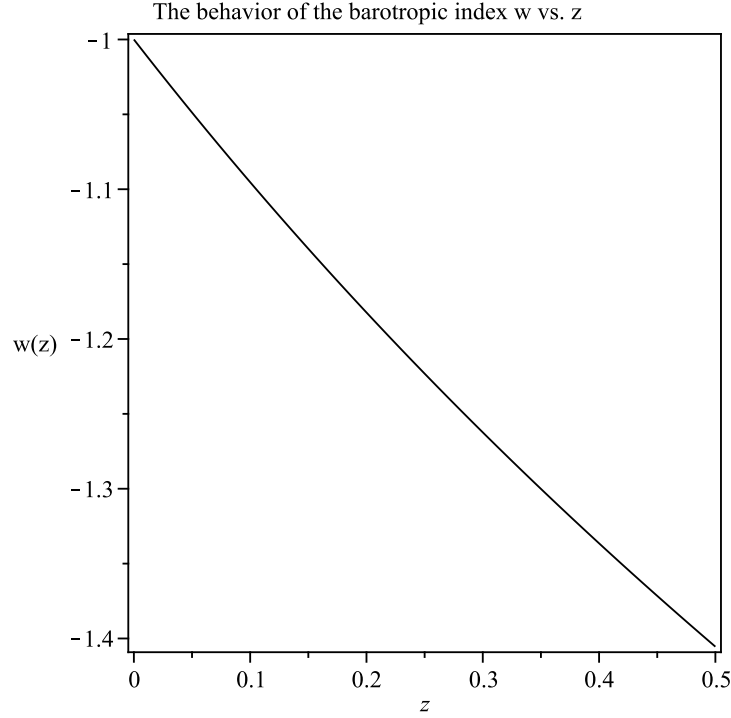} % scale goes from 0 to 1.
  \caption{The behavior of the barotopic index $w$ vs. $z$ }
  \label{fig9.eps}
\end{figure}
In this subsection we discuss the dynamics of the dark energy. We
know that in a flat FRW model with a perfect fluid of pressure $p$
and energy density $\rho$, the barotropic index $w(t)$ is defined by
\begin{eqnarray}
 w(t)=-1-\frac{\partial_t\ln(\rho(t))}{3H},
\end{eqnarray}
which can be can be derived from the continuity equation. Using
(9),(10) with (29) we obtain:
\begin{eqnarray}
 w(t)=-1-\frac{1}{3\left( -1+n \right)c_{{n}}}\,np{\chi}^{\,{\frac {n}{2(-1+n)}}}{t}^{\,{\frac
 {pn+2-2\,n}{2(-1+n)
}}}.
\end{eqnarray}

 For simplicity we take $D=2$,
 $\chi<<1$ and $G=\frac{1}{8\pi}$. Like the previous
sections, we are interested in expressing $w$ as a function of $z$.
 We substitute $\beta=\frac{2+(p-2)n}{2(1-n)}$ and from Eqs. (24) and (30) we
have
\begin{eqnarray}
 w(t)&=&-1-\frac{1}{3\left( -1+n \right)c_{{n}}}\,np{\chi}^{\,{\frac {n}{2(-1+n)}}}{\Big(\Big[
 \frac{-\log(1+z)}{\alpha}\Big]^{\frac{1}{\beta}}}\Big)^{\,{\frac
 {pn+2-2\,n}{2(-1+n)
}}}.\nonumber
\end{eqnarray}
Since $\frac{1}{\beta}\times\frac{pn+2-2n}{2(1-n)}=1$, therefore the
previous equation reads
\begin{eqnarray}
 w(t)&=&-1+\frac{1}{3\alpha\left( -1+n \right)c_{{n}}}\,np{\chi}^{\,{\frac {n}{2(1-n)}}}
\log(1+z),\nonumber
\end{eqnarray}
Now using the values of $\rho_n$, $b_n$, $c_n$ and $\Lambda_n$, we
can write
 \begin{eqnarray}
 w(z)=-1-d_{{n}}  \ln  \left( 1+z \right),
 \end{eqnarray}
where
\begin{eqnarray}
 d_n=\frac{2p}{\sqrt{3 }}(\frac{n^n\chi^{2-n}}{\Lambda_0
 (1-n)})^{\frac{1}{2(1-n)}}.
 \end{eqnarray}

Also we observe from the previous figures, the model parameters do
not depend on $p$ significantly and since the fractal parameter
$\chi<<1$ therefore we limited ourselves only to $p=1$. From (30) it
is obvious that the barotropic index crosses the dark energy limit
$w(t)=-1$ for  current era $z=0$ several times for different values
of $\chi$ and $n$. For instance, using typical values
$\chi=0.05,n=0.5,\Lambda_0=.01826$, (30) becomes
\begin{eqnarray}
w(t)=-1- \ln  \left( 1+z \right).
 \end{eqnarray}
 %%%%%

Figure-9 shows the behavior of the barotopic index $w$ against $z$.
The pressure may be written as
 \begin{eqnarray}
 p(z)=-p_n (\ln(1+z)+d_n
\ln(1+z)^{\frac{2-n}{n}}),
 \end{eqnarray}
where
\begin{eqnarray}
p_n=\rho_n\Big(-\frac{\chi}{\alpha}\Big)^{\frac{1}{1-n}}.
\end{eqnarray}

\subsection{ Consistency of the model $G(t)=\frac{G}{t}$ with
observational constraints}

From observational data, as it was shown in \cite{pitjeva}, there is
a secular decrease of the gravitational constant $G(t)$. This time
rate of change is of order $\frac{\dot{G}}{G}=(-5. 9\pm 4.4)\times
10^{-14} yr^{-1}$ which was obtained from planetary data analysis.
Thus, in our treatment on fractional cosmology focusing on the model
$G(t)=\frac{G}{t}$, we can accept it as a reasonable model in good
agreement with observational constraints \cite{obs}.

\section*{Final remarks}

This work was motivated by a more recent work on fractional-action
cosmology with a periodic weight function in the action
\cite{nabulsi} which we extended using a power-law weight function.
By doing so we obtained a varying gravitational coupling constant
and also a time varying cosmological constant. For both of these
`constants', we used different ansatz to analyze the behavior of
other parameters. We also modeled dark energy in this paradigm and
obtained relevant cosmological parameters including distance
parameters and statefinder parameters. The former helps in
determining the nature of dark energy while the latter one classify
different dark energy candidates.

\subsection*{Acknowledgments}
The authors would like to thank R. A. El-Nabulsi for useful comments
and valuable suggestions during this work. Also, we thank E. Pitjeva
for pointing the note on secular decrease of the gravitational
constant $G(t)$. M. Jamil would like to thank NUST for providing
support to present this work at the III Italian-Pakistani Workshop
on Relativistic Astrophysics in Italy.

\section*{References}
%%%%%%%%%%%%%%%%%%%%%%%%%%%%%%%%%%%%%%%%%%%

\smallskip

\end{document}